# Thermal Conductivity of Superconducting UPt$_3$ at Low Temperatures

M. J. Graf,* S.-K. Yip and J. A. Sauls

*Department of Physics & Astronomy, Northwestern University, Evanston, IL 60208*
version: December 5, 1995

*We study the thermal conductivity within the $E_{1g}$ and $E_{2u}$ models for superconductivity in UPt$_3$ and compare the theoretical results for electronic heat transport with recently measured results reported by Lussier, Ellman and Taillefer. The existing data down to $T/T_c \approx 0.1$ provides convincing evidence for the presence of both line and point nodes in the gap, but the data can be accounted for either by an $E_{1g}$ or $E_{2u}$ order parameter. We discuss the features of the pairing symmetry, Fermi surface, and excitation spectrum that are reflected in the thermal conductivity at very low temperatures. Significant differences between the $E_{1g}$ and $E_{2u}$ models are predicted to develop at excitation energies below the bandwidth of the impurity-induced Andreev bound states. The zero-temperature limit of the $\hat{c}$ axis thermal conductivity, $\lim_{T \to 0} \kappa_c/T$, is <u>universal</u> for the $E_{2u}$ model, but non-universal for the $E_{1g}$ model. Thus, impurity concentration studies at very low temperatures should differentiate between the nodal structures of the $E_{2u}$ and $E_{1g}$ models.*

PACS: 74.25.Fy, 74.70.Tx

## Introduction

Recent experiments of the thermal conductivity at low temperatures in UPt$_3$ have narrowed the range of possible symmetry states for the order parameter. Among several proposed order parameter models the two-dimensional orbital representations coupled to a symmetry breaking field appear to be very promising candidates.[1-4] For UPt$_3$, which has a hexagonal crystal structure ($D_{6h}$), phase diagram studies[5,6] and transport measurements[7-9] lead to either an even-parity, spin singlet E$_{1g}$ or an odd-parity, spin triplet E$_{2u}$ pairing state. In this paper we compare theoretical results for the electronic thermal conductivity of these two models with recent experimental data on single crystals of UPt$_3$.[10,11] We use the quasi-classical theory of superconductivity to calculate the thermal conductivity with the formulism derived in Ref. [12]. Our study focuses on the low temperature limit where the heat transport is dominated by electronic quasiparticles scattering off impurities.

In order to develop a theoretical model that accurately predicts the low temperature behavior of clean anisotropic superconductors we must determine the basis functions for the order parameter. These functions are the eigenfunctions of the linearized gap equation and are therefore determined by the microscopic pairing interaction. Since the pairing mechanism is uncertain for the heavy fermion superconductors we are forced to adopt a phenomenological approach in which calculations of the transport coefficients, based on models for the pairing basis functions, are compared with experiment. Fortunately, crystal symmetry provides significant constraints on the basis functions. The eigenfunctions of the linearized gap equation form irreducible representations of the point group. Thus, the ground state order parameter and excitation gap often exhibit nodal lines or points on the Fermi surface, which reflect the symmetry class of the order parameter.[13] This is the case for the 2D representations that have been proposed as models for the ground state of $UPt_3$. The basis functions for the pairing states that are often used in calculations are spherical harmonic functions or tight-binding functions with the transformation properties of the appropriate representation. However, it is worth emphasizing that the basis functions are not uniquely defined by symmetry. Thus, quantitative analysis is expected to require more detailed knowledge of the pairing functions than is contained in the simplest models for the basis functions. Our analysis focuses on the low temperature limit for two reasons: (i) the transport coefficients exhibit universal features in the limit $T \to 0$, which reflect the symmetry of the pairing state, and (ii) the low temperature transport coefficients are determined by lower dimensional regions of the Fermi surface near the nodes of the gap function. Thus, the complete Fermi surface and the detailed shape of the excitation gap are not required in order to accurately calculate the transport properties at low temperatures, $T \ll T_c$.

In Ref. [12] it was shown that certain eigenvalues of the thermal conductivity tensor are universal (i.e., independent of the scattering rate) below a characteristic temperature set by the bandwidth of impurity-induced Andreev bound states. Whether or not a universal limit develops at low temperatures depends sensitively on the nodal structure of the excitation gap. Both the $E_{1g}$ and $E_{2u}$ models for the order parameter lead to an excitation gap with a nodal line in the basal plane and point nodes along the $\pm\hat{c}$ directions.[14] The difference in the excitation spectrum for these two states is that the gap opens linearly for small angles away from the $\hat{c}$ direction for the $E_{1g}$ order parameter, but quadratically for the $E_{2u}$ order parameter. This difference is reflected in $\lim_{T\to 0} \kappa_c/T$, which is universal for the $E_{2u}$ model, but non-universal for the $E_{1g}$ model. A detailed description of these asymptotic limits is given in Ref. [12].

Recently Fledderjohan and Hirschfeld[15] examined the temperature dependence of the anisotropy ratio, $\kappa_\perp/\kappa_\parallel$, for the $E_{1g}$ and $E_{2u}$ models and concluded that the data of Lussier $et\ al.$[10] favored an $E_{1g}$ state with point nodes vanishing linearly near the poles. The authors qualified this conclusion because their calculations assumed a spherical Fermi surface and lowest order spherical harmonics for the pairing basis functions. Very recently, Norman and Hirschfeld[16] carried out numerical calculations of the electronic thermal conductivity using a Fermi surface for UPt$_3$ constructed from local density approximation (LDA) calculations, and several models for the pairing basis functions (e.g., tight-binding functions and Fermi-surface harmonics). These authors found considerable improvement between theory and experiment for the absolute magnitudes of $\kappa_{\perp,\parallel}/T$ compared to the results in Ref. [15] for both E$_{1g}$ and E$_{2u}$ models. They reported a good fit between theory and experiment for the E$_{1g}$ state, but they did not find an adequate fit in the E$_{2u}$ case. However, we note that the calculations reported in Ref. [16] do not describe the data of Lussier *et al.*[10] accurately in the low-temperature region, $T < 0.3 T_c$, for either pairing symmetry; the data of Lussier *et al.*[10] for $\kappa/T$ does not exhibit the saturation shown in the theoretical curves of Ref. [16], which assume a scattering rate of order $\Gamma_0 = 0.1 T_c$. This same observation was made by Lussier *et al.*[10] and led them to conclude that the low temperature data for $\kappa_b/T$ is more compatible with $\Gamma_0 \approx 0.01 T_c$ and a pairing symmetry of E$_{2u}$. However, all of these conclusions are based on calculations with model basis functions that do not accurately reproduce the low energy excitation spectrum.

### The E-Rep Models

In our comparison of theory with experiment we pursue a different approach than Norman and Hirschfeld.[16] Instead of examining the effects of the multi-sheet (LDA) Fermi surface on the heat current, we model the excitation spectrum by an excitation gap that opens at line (*ab* plane) and point (*c* axis) nodes on the Fermi surface, and by the Fermi surface properties in the vicinity of the nodes (i.e., the Fermi velocities near the nodes). The low-temperature behavior of the transport coefficients probes 'lower-dimensional' regions of the Fermi surface, where the excitation gap vanishes, and is less sensitive to the overall geometry of the Fermi surface. It is therefore admissible to parametrize the nodal regions of the gap, and attempt to fit the nodal parameters in order to achieve accurate low temperature limits for the thermal conductivity along the principal axes of the crystal. The advantage of this approach is that we can quantitatively determine the phase space contributing to the low temperature transport coefficients and then examine in more detail the effects of impurity scattering and order parameter symmetry on the heat current, without having to know the overall shape of the Fermi surface or basis functions.

We consider two models for the order parameter in the low-temperature phase of UPt$_3$ which belong to different representations of the $D_{6h}$ symmetry group:[1] (1) the even-parity E$_{1g}$ (hybrid-I) pairing state, and (2) the odd-parity E$_{2u}$ (hybrid-II) state,

$$(\text{E}_{1g}) \quad \Delta(\vec{p}_f) = \Delta_0(T)\, \mathcal{Y}_{E_{1g}}(\vec{p}_f)\,, \tag{1}$$

$$(\text{E}_{2u}) \quad \vec{\Delta}(\vec{p}_f) = \Delta_0(T)\, \mathcal{Y}_{E_{2u}}(\vec{p}_f)\, \hat{\mathbf{c}}\,, \tag{2}$$

where $\hat{\mathbf{c}}$ is the quantization axis along which the pairs have zero spin projection.[2]

---

[1] Since we are interested in the low-temperature phase of UPt$_3$ we neglect the small effect of the symmetry breaking field.

[2] In this analysis spin has no effect on the transport coefficients beyond the connection between the direction of the spin quantization axis and the nodal structure of the E$_{2u}$ basis functions.

The corresponding basis functions are

$$\mathcal{Y}_{E_{1g}}(\vec{p}_f) = y_0\, p_{fz}(p_{fx} + ip_{fy})\left(1 + a_2 p_{fz}^2 + a_4 p_{fz}^4\right), \quad (3)$$

$$\mathcal{Y}_{E_{2u}}(\vec{p}_f) = y_0\, p_{fz}(p_{fx} + ip_{fy})^2\left(1 + a_2 p_{fz}^2 + a_4 p_{fz}^4\right), \quad (4)$$

where $y_0$ is chosen to normalize the basis functions $\mathcal{Y}_\Gamma$ such that $\Delta_0$ is the maximal gap. The coefficients $a_2$ and $a_4$ determine the variation of the gap near the nodes. In the vicinity of the equatorial line ($p_{fz} = 0$), $|\Delta(\theta)| \sim \mu_{\text{line}} \Delta_0 |\Theta|$, with $\mu_{\text{line}} = y_0$, while near the poles ($p_{fx} = p_{fy} = 0$), $|\Delta(\theta)| \sim \mu_{\text{point}} \Delta_0 |\Theta|^n$ with $\mu_{\text{point}} = (1 + a_2 + a_4)\, y_0$, and $n = 1$ for $E_{1g}$ and $n = 2$ for $E_{2u}$.[3] This parametrization allows us to adjust independently the opening of the gap (slope or curvature) at the line and point nodes. For $a_2 = a_4 = 0$ the standard spherical harmonic basis functions are recovered. The crucial difference between the $E_{1g}$ and $E_{2u}$ states lies in the opening of the gap with angle $\Theta$ at the polar point nodes, as shown in Fig. 1.

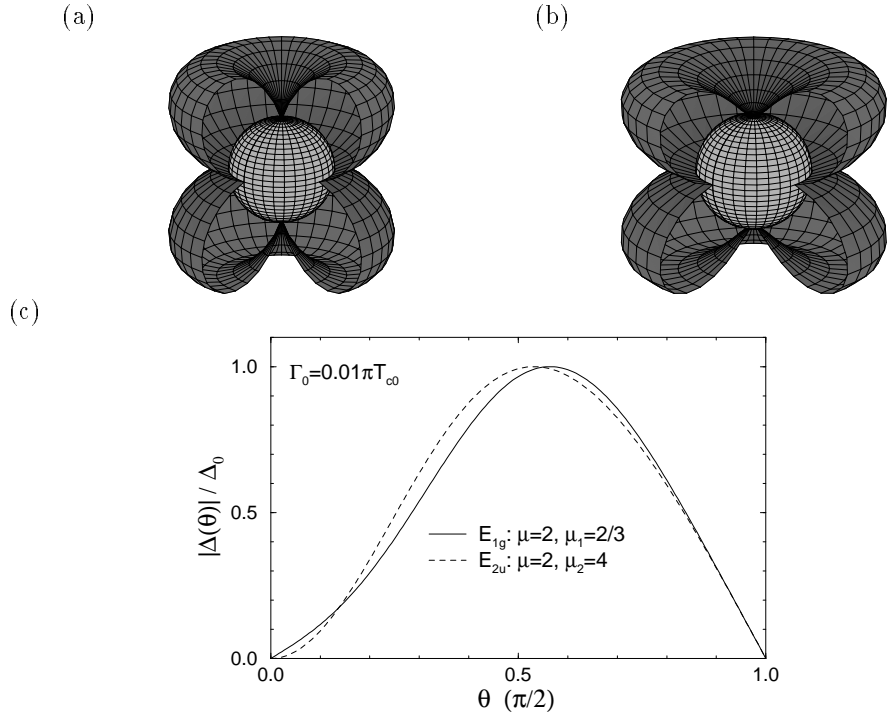

Fig. 1. The excitation gap (exaggerated) at the Fermi surface (sphere) for $E_{1g}$ (a) and $E_{2u}$ (b) order parameters showing the line node in the basal plane and the linear and quadratic point nodes. Panel (c) shows the normalized excitation gap at $T = 0$ for the same two states as a function of the polar angle $\Theta$. The parameters correspond to those that fit the low temperature thermal conductivity data shown in Figs. 2 and 3.

---

[3] The parameters $\mu_{\text{line}}$ and $\mu_{\text{point}}$ define the slope or curvature of the gap near a nodal line or point in a spherical coordinate system that is obtained by mapping an ellipsoidal Fermi surface onto a sphere.

The transport properties of unconventional superconductors are strongly influenced by scattering from impurities. One of the more striking effects is the appearance of a band of low energy excitations deep in the superconducting state.[17-19] This occurs for non-magnetic impurities in unconventional superconductors when the Fermi surface average of the order parameter vanishes, $\langle\Delta(\vec{p}_f)\rangle = 0$, i.e., the order parameter changes sign around the Fermi surface. Impurity bound states develop from the combined effects of impurity scattering and Andreev scattering. In an unconventional superconductor with a line of nodes a finite concentration of impurities leads to a band of Andreev bound states with a bandwidth, $\gamma < \Delta_0$, below which the density of states is almost constant and non-zero at zero energy.[20,21] This novel metallic band, deep in the superconducting phase, exhibits universal transport properties at very low temperatures.[12,22-25]

In Ref. [12] it was shown that the principal components of the thermal conductivity tensor can be expressed in terms of an *effective* transport scattering time that incorporates all of the coherence effects of superconductivity for $T \to 0$,

$$\lim_{T \to 0} \kappa_i(T) = \frac{v_{f,i}^2}{3} \gamma_S \, T \, \tau_i \quad (i = a, b, c). \tag{5}$$

Here $v_{f,i}^2$ is the Fermi-surface average of $[v_{f,i}(\vec{p}_f)]^2$, $\gamma_S = \frac{2}{3}\pi^2 k_B^2 N_f$ is the normal-state Sommerfeld coefficient, and the *effective* transport scattering time is defined by

$$\tau_i = \frac{3\hbar}{v_{f,i}^2} \int d\vec{p}_f \, \frac{[v_{f,i}(\vec{p}_f)]^2 \, \gamma^2}{[\Delta(\vec{p}_f)^2 + \gamma^2]^{\frac{3}{2}}} \,, \tag{6}$$

where $\gamma$ is the bandwidth of Andreev bound states. For heat flow in the basal plane one has $\tau_a = \tau_b = \tau_\Delta$, where $\tau_\Delta = 3\hbar/(4\mu_{\text{line}}\Delta_0(0))$, which is universal, i.e., independent of the scattering cross section or impurity concentration. The leading temperature corrections to $\kappa_b/T$ are given by a Sommerfeld expansion for $T \lesssim \gamma \ll T_c$,

$$\kappa_b(T) \simeq \alpha_b \, T + \beta_b \, T^3 \,. \tag{7}$$

In the strong scattering limit the parameters $\alpha_b$ and $\beta_b$ are related to the microscopic model parameters by

$$\alpha_b \simeq \frac{v_{f,b}^2}{3} \gamma_S \, \tau_\Delta \,, \qquad \beta_b/\alpha_b \simeq \frac{7\pi^2 k_B^2}{60 \, \gamma^2} \,. \tag{8}$$

For heat flow along the $\hat{\text{c}}$-axis the thermal conductivity again has a Sommerfeld expansion at low temperatures, $\kappa_c = \alpha_c T + \beta_c T^3$. However, the $\hat{\text{c}}$-axis coefficients depend sensitively on the winding number of the order parameter around the point nodes,[12] i.e., $n = 1$ for $E_{1g}$ and $n = 2$ for $E_{2u}$,

$$\alpha_c \simeq \frac{v_{f,c}^2}{3} \gamma_S \frac{\hbar}{2\,\mu_{\text{point}}\Delta_0(0)} \times \begin{cases} 2\,\gamma/\mu_{\text{point}}\Delta_0(0), & (E_{1g}) \\ 1, & (E_{2u}) \end{cases} \tag{9}$$

$$\beta_c/\alpha_c \simeq \frac{7\pi^2 k_B^2}{120\,\gamma^2} \times \begin{cases} 5, & (E_{1g}) \\ 2, & (E_{2u}) \end{cases}. \tag{10}$$

Note that $\alpha_c = \lim_{T \to 0} \kappa_c/T$ is universal for the $E_{2u}$ state, but depends on the impurity concentration $(n_i)$ for the $E_{1g}$ state. In the unitarity scattering limit $\gamma^2 \propto n_i$, thus, the coefficient of the $T^3$ term scales with the impurity concentration as $\beta_c \sim 1/n_i$ for the $E_{2u}$ state and $\beta_c \sim 1/\sqrt{n_i}$ for the $E_{1g}$ state.

### Numerical Results

We have carried out numerical calculations of the thermal conductivity, based on the formulism developed in Ref. [12]. These calculations coincide with the analytic results for the very low temperature limit described above. A principal input to the numerical calculations is the total quasiparticle scattering rate, $\Gamma(T)$, which is the sum of an elastic (impurity) and an inelastic (electron-electron) contribution (Matthiessen's rule). Estimates of the impurity scattering rate can be obtained from the residual resistivity or normal state thermal conductivity. The normal state scattering rate obtained from the thermal conductivity, $\Gamma(T) = \frac{1}{6} v_{f,b}^2 \, \gamma_S \, T/\kappa_b(T)$, is found to be nearly isotropic (s-wave scattering), and dominated by electron-electron scattering above the superconducting transition for clean single crystals of UPt$_3$.[10] In the temperature range, $0.5\,\text{K} \lesssim T \lesssim 1\,\text{K}$, the thermal conductivity is well described by the total scattering rate, $\Gamma(T) = \frac{\hbar}{2\tau(T)} = \Gamma_0 \left(1 + T^2/T_c^2\right)$;[10] for the particular crystal of Ref. [10] the elastic and inelastic contributions are nearly equal at $T_c$. The inelastic rate decreases rapidly at low temperature, even faster than $T^2$ because of the opening of a gap over most of the Fermi surface. However, below the superconducting transition we model the electron-electron scattering self-energy in the superconducting state by the same phenomenological, temperature dependent $\Gamma(T)$ as in the normal state. Thus, we neglect the reduction in $\Gamma(T)$ due to the onset of superconductivity. At $T = 0.1 T_c$ the inelastic rate is certainly less than 1% of the total scattering rate. Since our primary interest is in the low temperature behavior of the heat transport the error in the inelastic rate is insignificant. However, we do not expect as accurate a description of the experimental data in the temperature regime, $T_c/2 \lesssim T \leq T_c$, where inelastic scattering is not negligible.

In Figs. 2 and 3 we plot theoretical results for $[\kappa_i(T)\,T_c]/[\kappa_i(T_c)\,T]$ for the $E_{1g}$ (Fig. 2) and $E_{2u}$ (Fig. 3) models. Impurity scattering is assumed to be in the unitarity limit and the slope parameter of the line node is $\mu_\text{line} = 2$. The experimental data for UPt$_3$ shown in Figs. 2 and 3 was obtained from Ref. [11] and is normalized by the normal-state value $\kappa_b(T_c)/T_c \approx 20\,\text{mW}/(\text{K}^2\,\text{cm})$ at $T_c \simeq 0.5\,\text{K}$. The data is in excellent agreement with the theoretical curves for either symmetry class in the low temperature range, $0.1 T_c \lesssim T \lesssim 0.4 T_c$. The theoretical curves saturate for $T \lesssim T^\star \sim \gamma$ to the low temperature limits described by Eqs. (7)-(10). The experimental data shows no evidence of saturation down to the lowest temperature; thus, $T^\star < 0.1 T_c$. In addition to this constraint, the universal limit of the thermal conductivity in a superconductor with an equatorial line node implies that the ratio

$$\lim_{T \to 0} \frac{\kappa_b(T) T_c}{\kappa_b(T_c) T} \simeq \frac{3}{2} \frac{\Gamma(T_c)}{\mu_\text{line} \Delta_0(0)} \qquad (11)$$

scales linearly with the scattering rate, $\Gamma(T_c)$, and is independent of the scattering phase shift, $\delta_0$. The dependence on $\Gamma(T_c)$ in Eq. (11) comes from the normal-state value, $\kappa_b(T_c) \sim 1/\Gamma(T_c)$, and the parameter $\mu_{\text{line}} = (1/\Delta_0)|d\Delta(\theta)/d\theta|_{node}$ describes the opening of the gap at the line node.

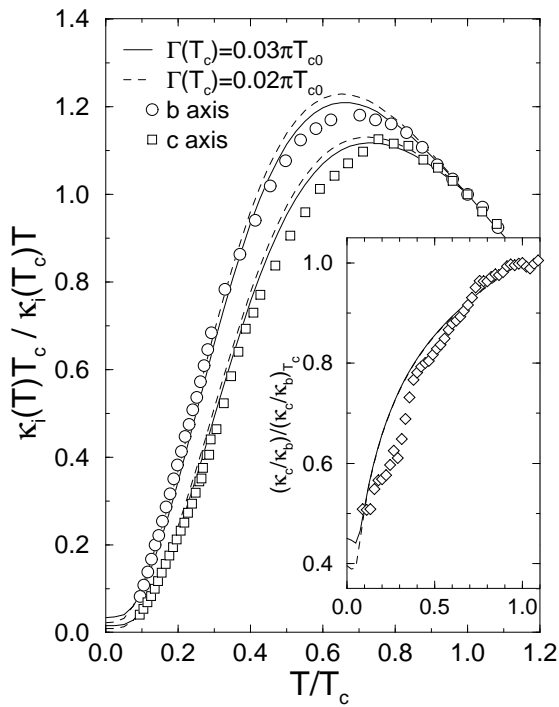

Fig. 2. The normalized thermal conductivity of an even-parity $E_{1g}$-state in the unitarity limit for heat flow along the $b$ and $c$ axes, and for two different scattering rates of $\Gamma(T_c) \equiv 2\Gamma_0$. The slope parameters at the linear point nodes and at the line node were fixed to $\mu_1 = 2/3$ and $\mu_l = 2.0$. The experimental data (symbols) were taken from Lussier et al.[11] and normalized to the value at $T_c \simeq 0.5\,\text{K}$. The inset shows the normalized anisotropy ratio $(\kappa_c/\kappa_b)/(\kappa_c/\kappa_b)_{T_c}$.

The value $\mu_{\text{line}} = 2$ is consistent with the constraints imposed by the intercept of $[\kappa_b(T)\,T_c]/[\kappa_b(T_c)\,T]$ at $T=0$ and the crossover temperature, $T^\star$, i.e., $\Gamma_0/(\mu_{\text{line}}\Delta_0) \lesssim \frac{1}{3}\cdot 0.08$ and $T^\star \simeq 0.2\,\sqrt{\Gamma_0\mu_{\text{line}}\Delta_0} \lesssim 0.1\,T_c$. The value of $\Gamma_0$ is adjusted to give the best fit to the low-$T$ region of the basal plane thermal conductivity. For $\Gamma_0 = 0.01\pi\,T_{c0} \approx 0.03\,T_c$ the intercept becomes, $[\kappa_b(T)/T]/[\kappa_b(T_c)/T_c]_{T=0} \approx 0.02$. This impurity scattering rate, that fits the low-temperature part of the thermal conductivity in the superconducting state, is smaller than that estimated from the normal-state transport and dHvA data, $\Gamma_0 \approx 0.1\,T_c - 0.2\,T_c$,[26] by at least a factor of 3. This discrepancy is considerably smaller than the discrepancy reported by several other authors.[11,10,15] Finally, the parameters defining the excitation gap near the point nodes are adjusted to fit $\kappa_c(T)/T$ at $T \ll T_c$. These values are $\mu_1 = 2/3$ for the $E_{1g}$ model and $\mu_2 = 4$ for the $E_{2u}$ model.

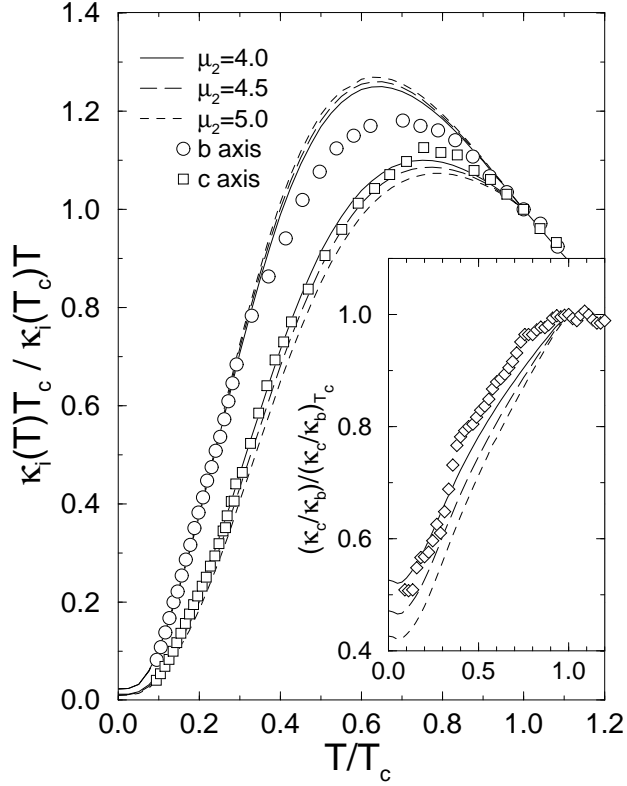

Fig. 3. The normalized thermal conductivity of an odd-parity $E_{2u}$-state in the unitarity limit for heat flow along the $b$ and $c$ axes, with a scattering rate of $\Gamma_0 = 0.01\pi T_{c0}$, and several choices for the curvature parameter at the point nodes, $\mu_2 = 4.0, 4.5, 5.0$. The slope of the line node was fixed to $\mu_{II} = 2.0$. The inset shows the normalized anisotropy ratio $(\kappa_c/\kappa_b)/(\kappa_c/\kappa_b)_{T_c}$.

Although the existing low temperature data for the thermal conductivity on UPt$_3$ is equally well described by either an $E_{1g}$ or $E_{2u}$ model, the ultra low temperature region offers a possibility of distinguishing these two models. The anisotropy ratio of the thermal conductivities $\kappa_c/\kappa_b$ as a function of temperature is shown in the insets of Figs. 2 and 3. The zero temperature limit of this ratio for the $E_{2u}$ (hybrid-II) state is a universal value, independent of impurity scattering,

$$\lim_{T \to 0} \frac{\kappa_c(T)/\kappa_b(T)}{\kappa_c(T_c)/\kappa_b(T_c)} \simeq \frac{\mu_{II}}{\mu_2} \quad (E_{2u}), \tag{12}$$

while this ratio in the case of the $E_{1g}$ (hybrid-I) state depends on the impurity concentration and scattering strength through the impurity bandwidth $\gamma$,

$$\lim_{T \to 0} \frac{\kappa_c(T)/\kappa_b(T)}{\kappa_c(T_c)/\kappa_b(T_c)} \simeq \frac{\mu_I \gamma}{\mu_1^2 \Delta_0(0)} \quad (E_{1g}). \tag{13}$$

In the unitarity limit $\gamma \sim \sqrt{\Gamma_0 \mu_{\text{line}} \Delta_0(0)}$. The slope and curvature parameters describe the opening of the order parameter at the line node ($\mu_{\text{I}}, \mu_{\text{II}}$) and at the point nodes ($\mu_1, \mu_2$). Thus, impurity concentration studies of the zero-temperature limit (i.e., $T < T^\star$) of the anisotropy ratio should distinguish between these two models for the ground state.[4]

The analysis has so far assumed unitarity scattering by impurities. It has been argued theoretically that impurity scattering in heavy fermion metals is generally expected to be in the unitarity limit.[27, 28] Fig. 4 shows the influence of the scattering phase shift $\delta_0$ on the thermal conductivity at low temperatures. It is clear that UPt$_3$ is in the strong scattering limit with a scattering phase shift $\delta_0 \gtrsim 80^\circ$. For strong scattering the crossover from the power law region to the universal limit occurs just below $T^\star \sim 0.1 T_c$. However, the crossover to the universal limit occurs at much lower temperatures for weak scattering, and is barely visible for the scattering phase shifts less than $\delta_0 = 60^\circ$.

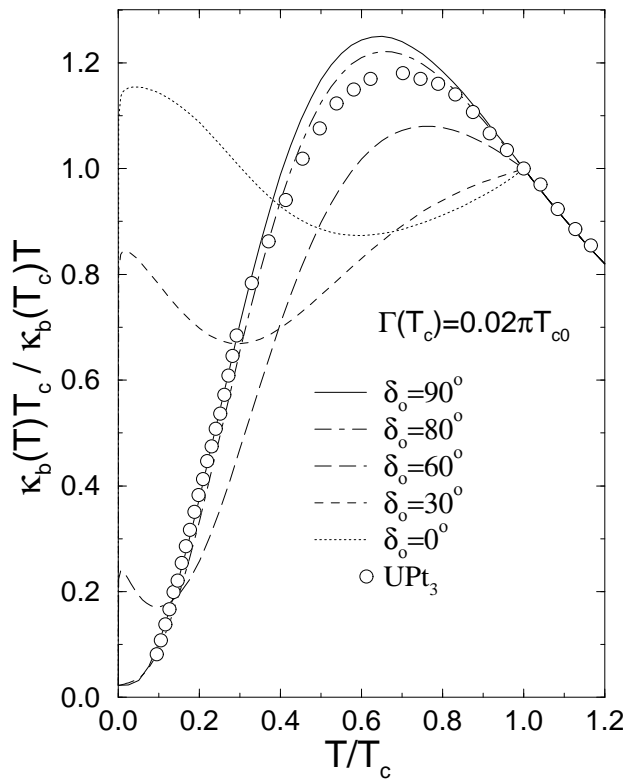

Fig. 4. The normalized thermal conductivity of the $E_{2u}$-state for heat flow in the basal plane (along the $b$-axis), a scattering rate $\Gamma(T_c) = 0.02\pi T_{c0}$, and various scattering phase shifts $\delta_0$. The slope and curvature parameters are $\mu_{\text{II}} = 2.0$ and $\mu_2 = 4.0$. For weak scattering, $\delta_0 \lesssim 60^\circ$, the crossover region to the universal behavior becomes very small, and barely observable in the figure. The experimental

---

[4]The same suggestion is made by Norman and Hirschfeld.[16]

data (circles) are from Lussier *et al.*[11]

The crossover to the universal limit is not seen in the currently available experimental data of Lussier *et al.*[11] The observation of saturation of $\kappa_i/T$ below $T^\star$ would be an important test of theory.[11,12] Recent measurements of heat transport in UPt$_3$ by Suderow *et al.*[29] and Huxley *et al.*[30] extend down to $T \approx 0.03\,T_c$. These measurements are in agreement with a crossover temperature $T^\star \approx 0.07\,T_c$. These authors also report an asymptotic value for the basal plane thermal conductivity of $[\kappa_a(T)/T]_{T\to 0} \approx 0.4\,\mathrm{mW/(K^2 cm)}$, which is in remarkably good agreement with our theoretical value, $[\kappa_b/T]_{T\to 0} \approx 0.02 \cdot 20\,\mathrm{mW/(K^2 cm)}$, extracted from the numerical fit to the data of Lussier *et al.*[10] shown in Figs. 2 and 3. A similar estimate for the universal value can be obtained with Eq. (8) from normal-state data. For UPt$_3$ it is reported that $\gamma_S \approx 10\,\mathrm{kJ/(K^2 m^3)}$[30] and $v_{f,b} \approx v_{f,c}/\sqrt{2.8} \approx 3.3\,\mathrm{km/s}$.[31,30] Thus, we estimate for the heat current in the basal plane $[\kappa_b(T)/T]_{T\to 0} \approx 1\,\mathrm{mW/(K^2 cm)}$, which is of the same order as our previous estimate[12] and is comparable to the value of $0.4\,\mathrm{mW/(K^2\,cm)}$ reported by Suderow *et al.*[29]

## Commentary

At low temperatures heat transport is dominated by elastic scattering of quasiparticles in low dimensional regions of the Fermi surface near nodes of the order parameter. As a result, the low energy spectrum can be determined accurately, without complete knowledge of the Fermi surface. Excellent agreement with experiment at low temperatures is obtained for both pairing states, E$_{1g}$ and E$_{2u}$. The differences in the excitation spectrum for these states are predicted to be observable at very low temperatures, $T < T^\star \sim 0.1\,T_c$, when the impurity-induced Andreev bound states determine the heat transport. Measurements of the thermal conductivity ratio, $\kappa_c/\kappa_b$, at very low temperatures for various impurity concentrations should differentiate between the two models; the E$_{2u}$ model is predicted to have a universal ratio, while this ratio for the E$_{1g}$ state is predicted to depend strongly on the concentration of impurities. At high temperatures, $T \sim T_c$, our calculations show deviations from the experimental data for both models, although the basal plane heat transport is in slightly better agreement for the E$_{1g}$ model with a scattering rate of $\Gamma(T_c) \approx 0.06\,T_c$. Accurate calculations of the heat transport in this temperature region require knowledge of the excitation spectrum over the entire Fermi surface, as well as a self-consistent treatment of inelastic scattering.

## Acknowledgments

We wish to thank L. Taillefer for stimulating discussions and sending us the data from his laboratory prior to publication, and the authors of Refs. [29,30] for sending us preprints of their work prior to publication. This research was supported by the National Science Foundation through the Science and Technology Center for Superconductivity (DMR 91-20000), and the Northwestern University Materials Science Center (DMR 91-20521).

# REFERENCES

[*] Electronic address: Graf@snowmass.phys.nwu.edu
1. H. v. Löhneysen, Physica B **197**, 551 (1994).
2. L. Taillefer, Hyp. Int. **85**, 379 (1994).
3. J. A. Sauls, Adv. Phys. **43**, 113 (1994).
4. R. Heffner and M. R. Norman, to appear in Comments Cond. Matt. Phys.; [http://xxx.lanl.gov/abs/cond-mat/9506043].
5. G. Bruls *et al.*, Phys. Rev. Lett. **65**, 2294 (1990).
6. S. Adenwalla *et al.*, Phys. Rev. Lett. **65**, 2298 (1990).
7. B. Shivaram, Y. Jeong, T. Rosenbaum, and D. Hinks, Phys. Rev. Lett. **56**, 1078 (1986).
8. C. Broholm, *et al.*, Phys. Rev. Lett. **65**, 2062 (1990).
9. P. J. C. Signore *et al.*, Phys. Rev. B **52**, 4446 (1995).
10. B. Lussier, B. Ellman, and L. Taillefer, Phys. Rev. Lett. **73**, 3294 (1995).
11. B. Lussier, B. Ellman, and L. Taillefer, to appear in Phys. Rev. B; [http://xxx.lanl.gov/abs/cond-mat/9504072].
12. M. J. Graf, S.-K. Yip, J. A. Sauls, and D. Rainer, preprint; [http://xxx.lanl.gov/abs/cond-mat/9509046].
13. G. E. Volovik and L. P. Gor'kov, Sov. Phys. JETP **61**, 843 (1985).
14. M. R. Norman, Physica C, **194**, 205 (1992).
15. A. Fledderjohann and P. J. Hirschfeld, Solid State Commun. **94**, 163 (1995).
16. M. R. Norman and P. J. Hirschfeld, preprint; [http://xxx.lanl.gov/abs/cond-mat/9509045].
17. L. J. Buchholtz and G. Zwicknagl, Phys. Rev. B **23**, 5788 (1981).
18. G. Preosti, H. Kim, and P. Muzikar, Phys. Rev. B **50**, 13 638 (1994).
19. A. V. Balatsky, M. L. Salkola, and A. Rosengren, Phys. Rev. B **51**, 15 547 (1995).
20. L. Gor'kov and P. Kalugin, JETP Lett. **41**, 253 (1985).
21. C. H. Choi and P. Muzikar, Phys. Rev. B **37**, 5947 (1988).
22. P. A. Lee, Phys. Rev. Lett. **71**, 1887 (1993).
23. M. J. Graf, M. Palumbo, D. Rainer, and J. A. Sauls, Phys. Rev B **52**, 10 588 (1995).
24. Y. Sun and K. Maki, Europhys. Lett. **32**, 355 (1995).
25. M. Palumbo and M. J. Graf, to appear in Phys. Rev B; [http://xxx.lanl.gov/abs/cond-mat/9508078].
26. L. Taillefer *et al.*, J. Magn. Magn. Mat. **63 & 64**, 372 (1987).
27. C. J. Pethick and D. Pines, Phys. Rev. Lett. **57**, 118 (1986).
28. S. Schmitt-Rink, K. Miyake and C. M. Varma, Phys. Rev. Lett. **57**, 2575 (1986).
29. H. Suderow, J. P. Brison, A. D. Huxley, and J. Flouquet, to appear in Physica B, SCES 1995 Proceedings.
30. A. D. Huxley, H. Suderow, J. P. Brison, and J. Flouquet, to appear in Phys. Lett. A (1995); J. P. Brison *et al.*, J. Low Temp. Phys. **95**, 145 (1994).
31. L. Taillefer and G. Lonzarich, Phys. Rev. Lett. **60**, 1570 (1988).